\documentclass[reprint,amsmath,amssymb,floatfix,onecolumn,superscriptaddress,aps,pre]{revtex4-2}
\usepackage{hyperref}
\usepackage{graphicx}
\usepackage{dcolumn}% Align table columns on decimal point
\usepackage{threeparttable}
\usepackage{bm}% bold math
\usepackage{color}
\usepackage{xcolor}
\usepackage{ulem}
\hypersetup{
    colorlinks=true,
    linkcolor=blue,
    citecolor=blue,   
   urlcolor=blue,
   }
\def\bea{\begin{eqnarray}}
\def\eea{\end{eqnarray}}
\def\beq{\begin{equation}}
\def\eeq{\end{equation}}

\def\be{\beta}

\def\t{\tau}

\def\la{\langle}
\def\ra{\rangle}
\def\nn{\nonumber}

\def\d{\delta}
\def\p{\partial}
\def\Pe{{\rm Pe}}

\def\uv{ \hat{\bm{u}}}
\def\rv{ {\bf r}}
\def\vv{ {\bf v}}

\def\d{\delta}
\def\p{\partial} 
\def\la{\langle}
\def\ra{\rangle}

\begin{document}

\title{Dynamical metastability and re-entrant localization of trapped active elements with speed and orientation fluctuations}

\author{Manish Patel}%
\email[]{manish.patel@iopb.res.in}
\affiliation{Institute of Physics, Sachivalaya Marg, Bhubaneswar-751005, Odisha, India}
\affiliation{Homi Bhabha National Institute, Anushakti Nagar, Mumbai 400094, India}

\author{Amir Shee}%
\email[]{amir.shee@northwestern.edu}
\affiliation{Northwestern Institute on Complex Systems and ESAM, Northwestern University, Evanston, IL 60208, USA}
\affiliation{Institute of Physics, Sachivalaya Marg, Bhubaneswar-751005, Odisha, India}

\author{Debasish Chaudhuri}
\email[]{debc@iopb.res.in}
\affiliation{Institute of Physics, Sachivalaya Marg, Bhubaneswar-751005, Odisha, India}
\affiliation{Homi Bhabha National Institute, Anushakti Nagar, Mumbai 400094, India}
\affiliation{Max-Planck Institute for the Physics of Complex Systems, N{\"o}thnitzer Strasse 38, 01187 Dresden, Germany}
\date{\today}%

\begin{abstract}
We explore the dynamics of active elements performing persistent random motion with fluctuating active speed and in the presence of translational noise in a $d$-dimensional harmonic trap, modeling active speed generation through an Ornstein-Uhlenbeck process. Our approach employs an exact analytic method based on the Fokker-Planck equation to compute time-dependent moments of any dynamical variable of interest across arbitrary dimensions.
We analyze dynamical crossovers in particle displacement before reaching the steady state, focusing on three key timescales: speed relaxation, persistence, and dynamical relaxation in the trap. Notably, for slow active speed relaxation, we observe an intermediate time metastable saturation in the mean-squared displacement before reaching the final steady state. 
The steady-state distributions of particle positions exhibit two types of non-Gaussian departures based on control parameters: bimodal distributions with negative excess kurtosis and heavy-tailed unimodal distributions with positive excess kurtosis.  
We obtain detailed steady-state phase diagrams using the exact calculation of excess kurtosis, identifying Gaussian and non-Gaussian regions and possible re-entrant transitions. 
\end{abstract}

\maketitle

\section{Introduction}
\label{sec_int}

Active systems consist of self-propelling entities that consume energy, breaking detailed balance and violating the equilibrium fluctuation-dissipation relation~\cite{Ramaswamy2010, Romanczuk2012, Marchetti2013, Elgeti2015, Bechinger2016, Baconnier2024}. They span various scales, from macroscopic phenomena like bird flocking~\cite{Reynolds1987, Nagy2010} and fish schooling~\cite{Partridge1980, Herbert-Read2011} to microscopic examples such as motor proteins~\cite{Kron1986, Ndlec1997}, bacteria~\cite{Keller1971, Berg1972, Berg1990, Diluzio2005, Lauga2006, Hill2007, Gachelin2014}, and cell motion~\cite{Selmeczi2008, Trepat2018, Henkes2020}.

Numerous artificial systems, like active colloidal microswimmers~\cite{Dhar2006, Howse2007, Erbe2008, Golestanian2009, Palacci2010}, active rollers~\cite{Bricard2013, Zhang2021}, and vibrated rods~\cite{Kumar2014}, have been developed to study active matter. The persistent motion of active agents is often modeled as active Brownian particles (ABPs), where particles maintain a constant active speed in a heading direction that diffuses randomly~\cite{Hagen2011, Sevilla2014, Kurzthaler2018, Basu2018, Malakar2018, Basu2019, Malakar2020, Santra2021, Shee2020, Chaudhuri2021, Caprini2022, Patel2023, SheePRE2022, patel2024exact, Pattanayak2024}. Although mostly neglected, the generation of active speed often involves stochastic internal processes leading to its fluctuation and relaxation.

The emergence of self-propulsion, whether through autocatalytic mechanisms in active colloids or intricate active processes in motile cells, is driven by internal stochastic dynamics that introduce inherent fluctuations in the active speed~\cite{Romanczuk2012, Schienbein1993, Schweitzer1998, Howse2007}. 
For example, in active colloids, self-propulsion is achieved through various stochastic physico-chemical mechanisms such as electrophoresis, thermophoresis, diffusiophoresis, and the Marangoni effect~\cite{kuron_18, jiang_10, MATSUDA201692}. These mechanisms generate local stochastic gradients in concentration, temperature, or surface tension, resulting in fluctuating active speeds.
Biological cells exhibit various kinds of taxis involving internal stochastic processes in response to different kinds of external cues, including chemotaxis (chemical signals), haptotaxis (surface-bound cues), durotaxis (mechanical changes), topotaxis (topographical features), galvanotactics (self-generated electric fields) etc.\cite{SenGupta2021}. These signaling pathways, involving the cell membrane and acto-myosin cytoskeleton, activate cellular protrusions and adhesion, enabling movement with fluctuating active speeds and orientations\cite{Bray_2001, SenGupta2021, Jain2020, Petrie2009}. Earlier studies have quantified active speed fluctuations in cells, such as in the galvanotactic motion of \textit{Human Granulocytes}, which exhibited an average speed of 
0.28 $\mu$m/s with fluctuations of 
$ \pm 0.15 \,\mu$m/s~\cite{Schienbein1993}. Similarly, \textit{Salmonella Typhimurium} showed a self-propulsion speed of 
$39.45\,\mu$m/s with fluctuations of  $\pm 6 \, \mu$m/s~\cite{otte_2021}. Despite these findings of speed fluctuations, most  studies, apart from some exceptions~\cite{Romanczuk2012, Schienbein1994, Peruani2007, otte_2021, Shee2022ActiveCrossovers}, focus on the persistent motion of active particles at constant speeds.

Further, confinement is another aspect that plays a significant role in determining properties of active systems~\cite{Charras2014, Paul2017, Winkler2019, Xi2017, Zanotelli2019, Ma2015, Mahmud2009}. External trapping provides insights into cell migration, sorting, and targeted drug delivery~\cite{Zhong2013, Minji2019, Geyer2013}. Cells naturally exist within the confines of extracellular matrices or neighboring cells, and confinement affects migration patterns~\cite{Winkler2019}. Different geometries lead to unique collective migration patterns~\cite{Xi2017, Zanotelli2019} and influence the formation of cardiac microchambers~\cite{Ma2015}.

Active Brownian particles demonstrate intriguing behavior, such as accumulation at confinement boundaries~\cite{Fily2015, Timo2024, Granek2024}. Recent experiments~\cite{Li2009, Takatori2016, Dauchot2019, Maggi2014} and theoretical studies~\cite{Das2018, Basu2019, Malakar2018, Dhar2019, Malakar2020, Wagner2017, Elgeti2015, Chaudhuri2021} have shown a transition in the steady-state distribution from a passive equilibrium-like Gaussian at the trap center to a distinctly active non-Gaussian distribution with off-center peaks, influenced by trap stiffness and active velocity~\cite{Malakar2020, Chaudhuri2021}.

In this study, we investigate active Brownian particles (ABPs) with fluctuating speed in a $d$-dimensional harmonic trap, using the Schienbein-Gruler mechanism for active speed generation~\cite{Schienbein1993, Schienbein1994, otte_2021, SheeJSTAT2022}. Our key contribution is demonstrating the exact time evolution of all dynamical moments in arbitrary dimensions through a Laplace transform of the governing Fokker-Planck equation. This approach, first introduced by Hermans et al. in 1952 for worm-like chains~\cite{Hermans1952}, has recently been applied to several models involving ABPs~\cite{Shee2020, Chaudhuri2021, SheePRE2022,SheeJSTAT2022,Pattanayak2024,Patel2023, patel2024exact}.
In this paper, we derive the exact expression for the time evolution of several position moments, which agrees well with numerical simulations. Position dynamics eventually reach a steady state over time, and we analyze dynamical crossovers, revealing two regimes for slow and fast active-speed relaxation. For slow speed relaxation under a strong trap, we find an intermediate-time quasi-steady trapping displayed by a valley in the MSD plot before eventual relaxation to the final steady state.

The active drive leads to departures from the equilibrium Gaussian behavior in the position probability distribution. We quantify this deviation by deriving an exact expression for the excess kurtosis in displacement, which depends on speed relaxation time and fluctuation strength. We present steady-state `phase diagrams' that illustrate non-Gaussian behavior as a function of speed relaxation time, speed fluctuation strength, activity, and trap strength. Note that this single-particle system can not exhibit a true phase transition; the phase diagram represents a crossover between active and passive behavior. The excess kurtosis shows both negative and positive departures from zero, indicating bimodal and heavy-tailed unimodal distributions, respectively. The negative values characterize ABP accumulation away from the trap center, while the positive values correspond to ABP clustering at the trap center in the presence of large displacement fluctuations having distribution tails longer than Gaussian. Non-Gaussian departures increase with activity; systems with fast speed relaxation and weak speed fluctuation exhibit negative departures, while those with slow relaxation and strong speed fluctuation show positive departures. Varying the trap strength reveals a re-entrant behavior from Gaussian to non-Gaussian and back. 

The paper is organized as follows: 
In Section~\ref{sec_model}, we present the model using Langevin dynamics, establishing a framework for calculating moments via the Laplace transform of the Fokker-Planck equation. Section~\ref{sec_firstmoments} details the calculation of the first moments, whereas Section~\ref{sec_secondmoments} examines the second moment of position, along with its dynamic crossovers and estimates of effective diffusion and pressure. Section~\ref{sec:fourthmoment} examines the fourth moment of position and presents phase diagrams based on excess kurtosis. Finally, Section~\ref{sec_conc} summarizes our main findings and presents an outlook.

\section{Model and Calculation of Moments}
\label{sec_model}
We consider an Active Brownian Particle (ABP) with fluctuating active speed in a $d$-dimensional harmonic trap of strength $k$. The overdamped dynamics are defined by the position $\rv' = (r'_1, r'_2, \ldots, r'_d)$ and active speed $v'$ along the orientation $\uv = (u_1, u_2, \ldots, u_d)$ over time $t'$. The active speed follows an Ornstein-Uhlenbeck process, relaxing to a mean speed $v'_{0}$ over a timescale $\t'_v$. The dynamics are influenced by Gaussian white noise with strengths $D$, $D_r$, and $D_v$ for translation, orientation, and active speed, respectively. We define $l = \sqrt{D/D_r}$ as the unit of length and $\t = 1/D_r$ as the unit of time, expressing all other quantities accordingly. Speed and velocity are measured in units of $\bar{v} = \sqrt{DD_r}$.
Using the dimensionless position $\rv = \rv'/l$, active speed $v = v'/\bar{v}$ and time $t = t'/\t$, we write the dimensionless form of Langevin dynamics in $d$ dimension in the following Ito form \cite{Ito1975}:
\bea
\label{eom1}d r_i &=& v u_i dt - \beta r_i dt + \sqrt{2}\, dB^{t}_{i}(t)\,, \\
\label{eom2}dv &=& - [(v - {\rm Pe} )/\t_v] dt + \sqrt{2A_v} \,dB^{s}(t)\,, \\
\label{eom3}du_i &=& (\d_{ij} - u_iu_j) \sqrt{2}\,dB^{r}_{j}(t) - (d-1) u_i dt\,.
\eea
In the above, we used the following dimensionless parameters: trap strength $\beta = \mu k \t$, mean activity $ {\rm Pe} = v_0/\bar{v}$, relaxation time $\t_v = \t'_v/\t$ and the strength of speed fluctuation $A_v = \t D_v/\bar{v}^2$. 
The Gaussian noise in translation, rotation, and speed relaxation dynamics are uncorrelated. 
The components of noise in translation and rotation obey $\la dB^{t}_i \ra=\la dB^{r}_i \ra = 0$ and $\la d B^{t}_i d B^{t}_j \ra = \la d B^{r}_i d B^{r}_j \ra = \d_{i,j}\, dt$, while for speed relaxation they obey $\la dB^s \ra = 0$ and $\la dB^s dB^s \ra = dt$.
Note that the first term in equation~(\ref{eom3}) projects the noise onto a ($d-1)$-\, dimensional surface, while the second term ensures the normalization of the unit vector after displacement.
The numerical integration of the Langevin equation can be performed using the direct Euler-Maruyama scheme.

Noting that the position $\rv$ and active speed $v$ perform drift-diffusion, while the orientation $\uv$ undergoes a diffusion on a $(d-1)$-dimensional hypersurface, we can write the corresponding Fokker-Planck equation for the probability distribution $P(\rv, v, \uv, t)$ as
\bea
\p_t P(\rv, v, \uv, t) &=& \nabla^2 P +  \nabla_u^2 P + A_v \partial_{v}^{2} P - v\, \uv\cdot \nabla P\nn\\
&&+\beta d P  +\beta\, \rv\cdot\nabla P +(1/\t_v) P+[(v-{\rm Pe})/\t_v] \partial_v  P\,, 
\label{eq:fp}
\eea
where $\nabla$ is the $d$-dimensional gradient operator and $\nabla_u^2$ is the spherical Laplacian in ($d-1$) dimensional orientation space. It can be expressed as $\nabla^2_u = z^2 {\sum_{i=1}^{d}} \p_{z_i}^2 - [z^2 \p_{z}^2 +(d-1) z \p_{z}]$ in the $d$-dimensional Cartesian coordinates with $u_i = z_i/z$ and  $z = |\bf{z}|.$ 

With the help of the Laplace transform of time $\tilde{P}(\rv, v, \uv, s) = \int_{0}^{\infty} dt\,e^{-st} P(\rv, v, \uv, t) $, the Fokker-Planck equation can be written as,
\bea
- P(\rv,v,\uv,0) + s \tilde{P}(\rv,v,\uv,s) &=& \nabla^2  \tilde{P} +  \nabla_u^2  \tilde{P} + A_v \partial_{v}^{2}  \tilde{P} - v\, \uv\cdot \nabla  \tilde{P}\nn\\
&& + {\nabla \cdot [ \be \rv \tilde{P}]}
%+\beta d  \tilde{P}  +\beta \rv\cdot\nabla  \tilde{P}
%
+(1/\t_v)  \tilde{P} +[(v-\Pe)/\t_v] \partial_v   \tilde{P}\,,
\label{eq:fp1}
\eea
where $P(\rv, v, \uv, 0)$ sets the initial condition.
This enables the formulation of the equation describing the mean value of any arbitrary observable in the Laplace domain defined as $\la \psi \ra_s = \int d\rv \,  dv\, d\uv\, \psi(\rv, v, \uv ) \tilde P(\rv, v, \uv, s)$. 
Multiplying the above equation by $\psi (\rv , v ,\uv)$ and integrating over all possible $(\rv, v, \uv)$, we get the moment equation
\bea
\label{eq:moment}
 -\la \psi \ra_0 + s \la \psi \ra_s &=& \la \nabla^2 \psi \ra_s + \la \nabla_u^2 \psi \ra_s + A_v \la \p_v^2 \psi \ra_s +  \la v \uv \cdot \nabla \psi \ra_s - \beta \la \rv \cdot \nabla \psi \ra_s \nn\\
 &&- (1/\t_v) \la (v -{\rm Pe})  \p_v \psi \ra_s\,, 
\eea
where  $\la \psi \ra_0 = \int d\rv \, dv\, d\uv\,  \psi(\rv, v, \uv) P(\rv, v, \uv, 0)$. Without any loss of generality, the initial condition can be chosen as $P(\rv,v,\uv,0) = \d(\rv - \rv_0) \d(v-v_1) \d(\uv - \uv_0)$.
This represents an initial state where the system is precisely located at position $\rv_0$, moving at speed $v_1$, and oriented in the direction $\uv_0$. 
We utilize equation~(\ref{eq:moment}) to derive the exact form of the moments as a function of time. For example, the orientation autocorrelation obtained using equation~(\ref{eq:moment}) gives the expected result $\la \uv(t)\cdot\uv(0)\ra=e^{-(d-1)t}$ for persistent random walker. %~\cite{Shee2020}.
In the following, we calculate several dynamic moments and investigate the intriguing steady-state results using $\lim_{t \to \infty} \la \psi \ra = \lim_{s \to 0} s \la \psi \ra_s = \la \psi \ra_{\rm st}$.

%\section{Fokker-Planck equation}
%\label{sec:F-P}
\section{First Moments: Mean position and active velocity}
\label{sec_firstmoments}
We begin by exploring the time evolution of first moments, such as mean position and active velocity, using equation~(\ref{eq:moment}). Both the dynamics evolve to reach a steady state in the presence of a harmonic trap.

Using $\psi = \rv$ in equation~(\ref{eq:moment}) with the relations $\nabla^2 \rv = 0$, $\nabla_u^2 \rv = 0$, $\p_v^2 \rv = 0$, $v \uv \cdot \nabla \rv = v \uv$, $\rv \cdot \nabla \rv = \rv$ and $\p_v \rv = 0$, we get the mean position in the Laplace space $\la \rv \ra_s = [ \rv_0 + \la v \uv \ra_s]/(s+\beta)$. Again using $\psi = \boldsymbol{v} = v \uv$ in equation~(\ref{eq:moment}), we get the mean active velocity $\la \boldsymbol{v} \ra_s = [v_1 \uv_0 + \Pe\, \la \uv\ra_s/\t_v ]/(s+d-1+1/\t_v)$. To complete the calculation, we set $\psi = \uv$ in equation~(\ref{eq:moment}) to get $\la \uv \ra_s = \uv_0 /(s+d-1)$. Thus, we get the mean position and active velocity in the Laplace space as 
\bea
\label{eq:vels} \la \boldsymbol{v} \ra_s &=& \frac{(v_1 - \Pe) \uv_0}{(s+(d-1) + 1/\t_v)} + \frac{\Pe \, \uv_0}{(s+(d-1))}\,, \\
\label{eq:rs}
\la \rv \ra_s &=& \frac{(v_1 - \Pe) \uv_0}{(s+\beta)(s+d-1+\t_v^{-1})} + \frac{\Pe \, \uv_0 + \rv_0 (s+(d-1))}{(s+\beta) (s+(d-1))}\,.
\eea
The inverse Laplace transform of equation~(\ref{eq:vels}) and (\ref{eq:rs}) gives the full-time evolution as 
\bea
&\la \boldsymbol{v} \ra (t) = (v_1 - \Pe) \uv_0\, e^{-((d-1)+\t_v^{-1})t} + \Pe \, \uv_0 \, e^{-(d-1)t}\,, \\
\label{eq:ravg}
&\la \rv \ra (t) = \rv_0 e^{-\beta t} + \frac{(v_1 - \Pe) \uv_0}{(d-1)-\beta+\t_v^{-1}}\left( e^{-\beta t} - e^{-((d-1)+1/\t_v)t}\right) 
+\frac{\Pe\, \uv_0}{(d-1)-\beta} \left( e^{-\beta t} - e^{-(d-1)t} \right)\,. \nn\\
\eea
The mean position and active velocity show a $\t_v$ dependency only if $v_1 \neq \Pe$ and, as expected, vanish at the steady state. 
In the long time limit, the mean active speed approaches a nonzero steady-state value $\Pe$ (see Appendix~\ref{appendix:mean-speed} for further details).
 In the limit of slow relaxation $\t_v \ll 1$, the expression of mean position and active velocity reduces to the ABP with constant active speed in the presence of a harmonic trap, as studied in Ref.~\cite{Chaudhuri2021}.
Additionally, for $\beta = 0$, the dynamics recover that of free ABPs as described in Ref.~\cite{Shee2020}.

\section{Second Moments: Mean-Squared Displacement and Dynamical Crossover}
\label{sec_secondmoments}
In this section, we present the time evolution of the second moment of position and position-orientation cross-correlation.
We employ the analytical expression of mean-squared displacement (MSD) to describe the crossovers appearing in the dynamics.
The second moment of active velocity is identical to the second moment of active speed shown in Appendix~\ref{appendix:mean-speed}. 

% === Position-orientation cross-correlation
\subsection{Position-orientation cross-correlation}
The projection of position along the orientation is the simplest second moment to begin with.
Following similar steps as before, we set $\psi=\uv \cdot \rv$ in equation~(\ref{eq:moment}) to obtain the position-orientation cross-correlation in the Laplace space as
\bea
\la \uv \cdot \rv \ra_s = \frac{v_1-\Pe} {[s+(d-1)+\beta](s+\t_v^{-1})} + \frac{\Pe + s\, \uv_0\cdot \rv_0}{s[s+(d-1)+\beta]}\,.
\label{eq:cross_corr_Laplace} 
\eea
The inverse Laplace transform of the above equation gives the complete time evolution 
\bea
 \la \uv \cdot \rv \ra (t) &=& \frac{v_1-\Pe}{(d-1)+\beta -\t_v^{-1}} \left( e^{-t/\t_v} - e^{-((d-1)+\beta)t} \right) + \frac{\Pe}{(d-1)+\beta} \left( 1 - e^{-((d-1)+\beta)t}\right) \nn\\
&&+ (\uv_0 \cdot \rv_0) e^{-((d-1)+\beta)t}\,. 
\label{eq:cross_corr}
\eea
Similar to the first moments, the impact of $\tau_v$ on $\la \uv \cdot \rv \ra (t)$ can be seen only if $v_1 \neq \Pe$.
In the limit of fast relaxation with $\t_v \ll 1$, equation~(\ref{eq:cross_corr}) restores the result of constant speed ABPs in a harmonic trap \cite{Chaudhuri2021}, while in the other limit of vanishing trap stiffness $\beta \ll 1$, it restores the result of free ABPs with speed fluctuation~\cite{SheeJSTAT2022}.
Additionally, in the limit of both $\beta \ll 1$ and $\t_v \ll 1$, it restores the result of free ABPs \cite{Shee2020}.

 In the steady-state limit, equation~(\ref{eq:cross_corr}) saturates to $\la \uv \cdot \rv \ra_{st}=\Pe/[(d-1)+\beta]$, independent of $\t_v$. 
 %\alert
 {Thus the dimensionless swim pressure at steady state ${\cal P} = \rho (\Pe/d) \la \uv \cdot \rv \ra_{st}$~\cite{Omar2020, Bialke2015, Takatori2014, Mallory2013a} can be expressed as
\bea
{\cal P} = \rho \frac{\Pe^2}{d(d-1+\beta)}\,,
\eea
independent of the active speed relaxation and fluctuations, {%\color{blue} 
where $\rho$ is the dimensionless particle density}. However, this distinguishes between pressure generated at shallow or stiff potential. Clearly, pressure for stiff potential (larger $\be$) is low, in agreement with the notion of high (low) pressure at soft (stiff) boundaries due to ABP gas. This property was used before to demonstrate the remarkable displacement of a mobile partition with stiff and soft sides immersed in the middle of an ABP gas~\cite{Solon2015}, showing an absence of equation of state.}

%%%%%%%%%%%%%
%%%%%%%%%%%%%
% === Mean-squared displacement ===
\subsection{Mean-squared displacement}
\label{MSD} 
To compute MSD, we begin by setting $\psi=\rv^2$ in equation~(\ref{eq:moment}). 
It can be readily observed that $\la \psi \ra_0 = \rv_{0}^{2}$, $\la \nabla_u^2 \psi \ra_s =0$, $\la \p_v^2 \psi \ra_s = 0$, and $\la (v-\Pe) \p_v \psi \ra_s = 0$. 
The remaining averages are $\la \nabla^2 \psi \ra_s = 2d \la 1 \ra_s = 2d/s$ \footnote{As, $\la 1 \ra_s = \int {\rm d}\rv\,{\rm d}v\,{\rm d}\uv \tilde{P} =\int {\rm d}\rv\,{\rm d}v\,{\rm d}\uv \int_{0}^{\infty} {\rm d}t e^{-st} P = \int_{0}^{\infty} {\rm d}t e^{-st} \{\int {\rm d}\rv\,{\rm d}v\,{\rm d}\uv P\} = \int_{0}^{\infty} {\rm d}t e^{-st} = 1/s $}, $\la v\, \uv \cdot \nabla \psi \ra_s = 2 \la v \,\uv \cdot \rv \ra_s $, and $\la \rv \cdot \nabla \psi \ra_s = 2 \la \rv^2 \ra_s$.
Substituting these values into equation~(\ref{eq:moment}) we get
\bea
\la \rv^2 \ra_s = \frac{1}{(s+2\beta)} \left[ \rv_0^2 + \frac{2d}{s} + 2 \la v\, \uv \cdot \rv \ra_s  \right]\,, \nn
\eea
where $\la \rv^2 \ra_s$ depends on $\la v \, \uv \cdot \rv \ra_s$. 
To evaluate $\la v \, \uv \cdot \rv \ra_s$, we again set $\psi = v\, \uv \cdot \rv$ in equation~(\ref{eq:moment}) to get
  \bea
  \la v \,\uv \cdot \rv \ra_s = \frac{[v_1\,\uv_0 \cdot \rv_0 + \la v^2 \ra_s + \Pe \la \uv \cdot \rv \ra_s/\t_v]}{(s+(d-1)+\beta +\t_v^{-1})}\,. \nn
  \eea
Thus, plugging the above relation in the expression of $\la \rv^2\ra_s$ one finds
\bea
\la \rv^2 \ra_s = \frac{s \rv_0^2 +2d}{s(s+2 \beta)} + \frac{2[v_1\,\uv_0 \cdot \rv_0 + \la v^2 \ra_s + \Pe \la \uv \cdot \rv \ra_s/\t_v]}{(s+2\beta) (s+(d-1)+\beta +\t_v^{-1})}\,, 
\label{eq:r2ds}
\eea
where $\la \uv \cdot \rv \ra_s$ is given by equation~(\ref{eq:cross_corr_Laplace}). To further complete the calculation for $\la \rv^2 \ra_s$, we use equation~(\ref{eq:moment}) in the similar fashion to get $\la v^2 \ra_s = [v_1^2 + 2 A_v /s + 2\Pe \la v \ra_s/\t_v]/(s+2/\t_v) $, and $\la v \ra_s = (v_1 + \Pe/(s \t_v))/(s+1/\t_v) $. 
Putting these back into equation~(\ref{eq:r2ds}), we get the full form of $\la \rv^2 \ra_s$ in the Laplace space as
\begin{align}
\label{eq:r2sfull}
 \la \rv^2 \ra_s &= \frac{1}{s+2 \be} \left[  \rv_0^2 + \frac{2d}{s} + \frac{2}{s+(d-1)+\be + \t_v^{-1}} \left( \frac{2 \Pe^2 + 2 s\Pe \t_v v_1 +  \t_v (1+s\t_v) (2 A_v + s v_1^2)}{s(s \t_v + 1)(s \t_v + 2)}   \right. \right. \nn\\
&\left. \left. +v_1 \uv_0 \cdot \rv_0 + \frac{\Pe (\Pe + s(\uv_0 \cdot \rv_0 (1 + s \t_v) + \t_v v_1))}{s \t_v(s + d-1+\be)(s \t_v + 1)}   \right) \right]\,.
\end{align}
The inverse Laplace transform of equation~(\ref{eq:r2sfull}) provides the full-time evolution of MSD for a given initial state. However, since the expression is too lengthy, we show the evolution by plotting it in Fig.~\ref{fig:fig1} and analyzing its properties in Appendix~\ref{appendix:msdarb}.
However, for the specific choice of initial conditions $\rv_0 = \mathbf{0}$ and $v_1 = \Pe$, the expression of MSD and higher moments simplify significantly, making them easier to discuss. In the subsequent sections, we adopt these initial conditions for position and active speed to present and analyze all results. For these initial conditions, $\rv_0 = \mathbf{0}$ and $v_1 = \Pe$, the MSD is given by
\begin{align}
 \la \rv^2 \ra (t) &= \left( \frac{d}{\beta} + \frac{\Pe^2}{\beta(d-1-\beta)} \right) \left(1- e^{-2\beta t} \right) - \frac{2 \Pe^2}{((d-1)^2 -\beta^2)} \left( 1- e^{-(d-1+\beta)t} \right) \nn\\
  &+ \frac{A_v}{(d-1+\beta -\tau_v^{-1})(\beta - \tau_v^{-1})}\left( \tau_v (1-e^{-2t/\tau_v}) - \frac{(1-e^{-2\beta t})}{\beta}  \right) \nn\\
 &+ \frac{4 A_v}{[(d-1)^2 - (\beta - \tau_v^{-1})^2]} \left( \frac{1-e^{-(d-1+\beta + 1/\tau_v)t}}{(d-1+\beta+\tau_v^{-1})} - \frac{(1-e^{-2\beta t})}{2\beta} \right)\,.
\label{eq:r2d2}
\end{align}
The plot of the above expression captures the numerical simulation result of MSD, as shown in figure~\ref{fig:fig1}($a$) and ($b$). Notably, unlike $\la \rv \ra (t)$ and $\la \uv \cdot \rv \ra (t)$, $\tau_v$ impacts the time evolution of $\la \rv^2 \ra (t)$ even when $v_1 = \Pe$.
In the absence of speed fluctuations $A_v = 0$, the initial choice of active speed $v_1 = \Pe$ ensures that the active speed remains constant at $v = \Pe$ for equation~(\ref{eq:r2d2}), thereby restoring the result for constant speed ABPs in the harmonic trap \cite{Chaudhuri2021}. 
Furthermore, in the absence of both harmonic trap and speed fluctuations, MSD in the equation~(\ref{eq:r2d2}) restores the result for free ABPs \cite{Shee2020}.

In the asymptotic long time limit, MSD saturates to the steady state value
\bea
\la \rv^2 \ra_{\rm st} = \frac{d}{\beta} + \frac{\Pe^2}{\beta (d-1+\beta)} + \frac{\tau_v A_v}{\beta[ (d-1+\beta) + \t_v^{-1}]}\,, 
\label{eq:r2d_steady_state}
\eea
which depends on fluctuation strength and the two time scales $\t_v$ and $\be^{-1}$.
The steady-state MSD increases monotonically with $\t_v$ and $A_v$. It diverges for very large values of either parameter. With $\beta$, it monotonically decreases and vanishes as $1/\beta^2$ in the large $\beta$ limit. 
\subsection{Effective Diffusion}
The fluctuation in position at equilibrium satisfies $\la \rv^2_{\rm st} \ra = d D/\beta D_r$, where $D = k_{B}T/\gamma$ is defined by fluctuation-dissipation relation (FDR).
Away from equilibrium, the FDR no longer holds in its conventional form. However, by extending the equilibrium notion, we can obtain an estimate of the effective diffusivity
\bea
D_{\rm eff} := \frac{\beta \la \rv^2 \ra_{\rm st}}{d} = 1 + \frac{\Pe^2}{d(d-1+\beta)} + \frac{\t_v\, A_v}{d[ (d-1 + \beta )+ \t_v^{-1}]}\,,
\eea
which underscores the critical role of $\t_v$, $A_v$ and interaction $\beta$. Clearly, the effective diffusivity is an increasing function of the speed relaxation time $\t_v$ and speed fluctuation $ A_v$.
A sanity check can be obtained in the free particle limit of  $\beta \to 0$, in which the diffusivity reduces to $D_{\rm eff}=1+ \Pe^2/d(d-1) + \t_v A_v/d[(d-1)+\t_v^{-1}]$, a result established before for free ABPs with speed fluctuation~\cite{SheeJSTAT2022}.

\subsection{Dynamical Crossovers and Probability Distribution}
Here, we analyze the series expansion of $\la \rv^2 \ra (t)$ to identify the crossovers in the dynamics and to relate them to the dynamical change in probability distribution as obtained from direct numerical simulations. 
The nature and occurrence of these crossovers highly depend on the timescales $\t_v$ and $\be^{-1}$.

The series expansion of MSD with initial conditions $v_1=\Pe$ and $\rv = \bf{0}$ from equation~(\ref{eq:r2d2}) around $t = 0$ gives 
\bea
 \la \rv^2 \ra (t) = 2d\,t+ (\Pe^2 - 2d \beta)t^2 + \frac{1}{3} \left( 2 A_v + 4d \beta^2 - (d-1+3 \beta) \Pe^2 \right) t^3 + {\cal O}(t^4)\,.
\label{eq:r2d_small_time}
\eea
 The first term in the expansion suggests a diffusive scaling at a short time $\la\rv^2\ra\sim t$, set by thermal diffusivity. It then crosses over to a ballistic scaling  $\la\rv^2\ra\sim t^2$ at time $t_{I}=2d/(\Pe^2 -2d\beta)$ if $\Pe^2 > 2d \beta$.

{%\color{violet} 
Beyond this, further crossovers in MSD show qualitative differences depending on the competition of two timescales $\t_v$ and $\be^{-1}$, which results in four possibilities at intermediate time $t$: (a)~$t \ll (\t_v, \be^{-1})$,  (b)~$\be^{-1} \ll t \ll \t_v$,  (c)~$ \t_v \ll \t \ll \be^{-1}$ and  (d)~$ (\t_v,\be^{-1}) \ll t$. 
In the first regime of slow speed relaxation and weak trap, $t \ll (\t_v, \be^{-1})$, the system can spend a significantly long time in the initial diffusive regime before crossing over to ballistic behavior in the intermediate time in a manner described by equation~(\ref{eq:r2d_small_time}).  In the subsequent sections, we explore the intermediate time dynamics for the remaining three cases in further detail.

\subsection*{(i) Strong trap strength and slow speed relaxation, $\beta^{-1} \ll t \ll \t_v$, shows metastability:}
Here, MSD exhibits an intermediate time crossover to a quasi-steady state, as shown in figure~\ref{fig:fig1}($a$).
This transition occurs as the strong trap influences ABPs to relax to it more quickly before active speed can relax to the steady state value.
The nature of this crossover at an intermediate time $\beta^{-1} \ll t \ll \t_v$ can be obtained explicitly  using $e^{-2 \be t} \approx 0$, $e^{-(d-1+\be + 1/\t_v)t} \approx 0$, $e^{-(d-1+\be)t} \approx 0$ and $e^{-2t/\t_v} \approx 1 - 2t/\t_v$ in equation~(\ref{eq:r2d2}).
Using these we get the following form of MSD: $ \la \rv^2 \ra_{\rm int} = {\cal A} + {\cal B} t $, where
\begin{align}
{\cal A} &= \frac{d}{\beta} + \frac{\Pe^2}{\beta (d-1+\beta)} + \frac{A_v [(d-1 +3\beta) -\t_v^{-1}] }{\beta (\beta -\t_v^{-1}) [ (d-1+\beta)^2 -\t_v^{-2}]}\,,\\
 {\cal B} &=  \frac{2 A_v }{(\beta  - \t_v^{-1})[(d-1+\beta) - \t_v^{-1}]}\,.
\end{align}
As noted before, for $\Pe^2 > 2d\beta$, MSD cross overs from initial thermal diffusion $\la \rv^2 \ra \sim t$ to ballistic scaling $\la \rv^2 \ra \sim t^2$ at $t_{I}$. In a strong trap, the ABP with slow speed relaxation shows an intermediate time crossover from ballistic to a quasi-steady state $\la \rv^2 \ra \sim t^0$ at $t_{II} = [{\cal A}/(\Pe^2 - 2d \beta)]^{1/2}$, which is estimated by comparing the second term in the right-hand side of equation~(\ref{eq:r2d_small_time}) with the  $\mathcal{A}$ term in the expression of $ \la \rv^2 \ra_{\rm int} $.
Further, the quasi-steady state behavior crosses over to a subsequent diffusive scaling at $t_{III} = {\cal A}/{\cal B}$ provided $t_{III} > t_{II}$. 
At late time, this diffusive scaling crosses over to the final steady state at $t_{IV} = \la \rv^2 \ra_{\rm st}/{\cal B}$, estimated by the comparison of the second term in the right-hand side of $\la \rv^2 \ra_{\rm int}$ with the equation\,(\ref{eq:r2d_steady_state}).

Note that, for small activity $\Pe^2 < 2d \beta$, the MSD directly crosses over from thermal diffusive scaling at the shortest time to the quasi-steady state at $t_{I}^{'} = {\cal A}/2d$, obtained by comparing the first term in the right-hand side of equation\,(\ref{eq:r2d_small_time}) with the first term in the right-hand side of $\la \rv^2 \ra_{\rm int}$.
This quasi-steady state crosses over to subsequent diffusive scaling at $t_{III}$ obtained above. As before, in the asymptotic long-time, the diffusive scaling crosses over to the final steady state at $t_{IV}$.

% === FIGURE 1 ===

\begin{figure}
    %\centering
    \includegraphics[width = 20 cm]{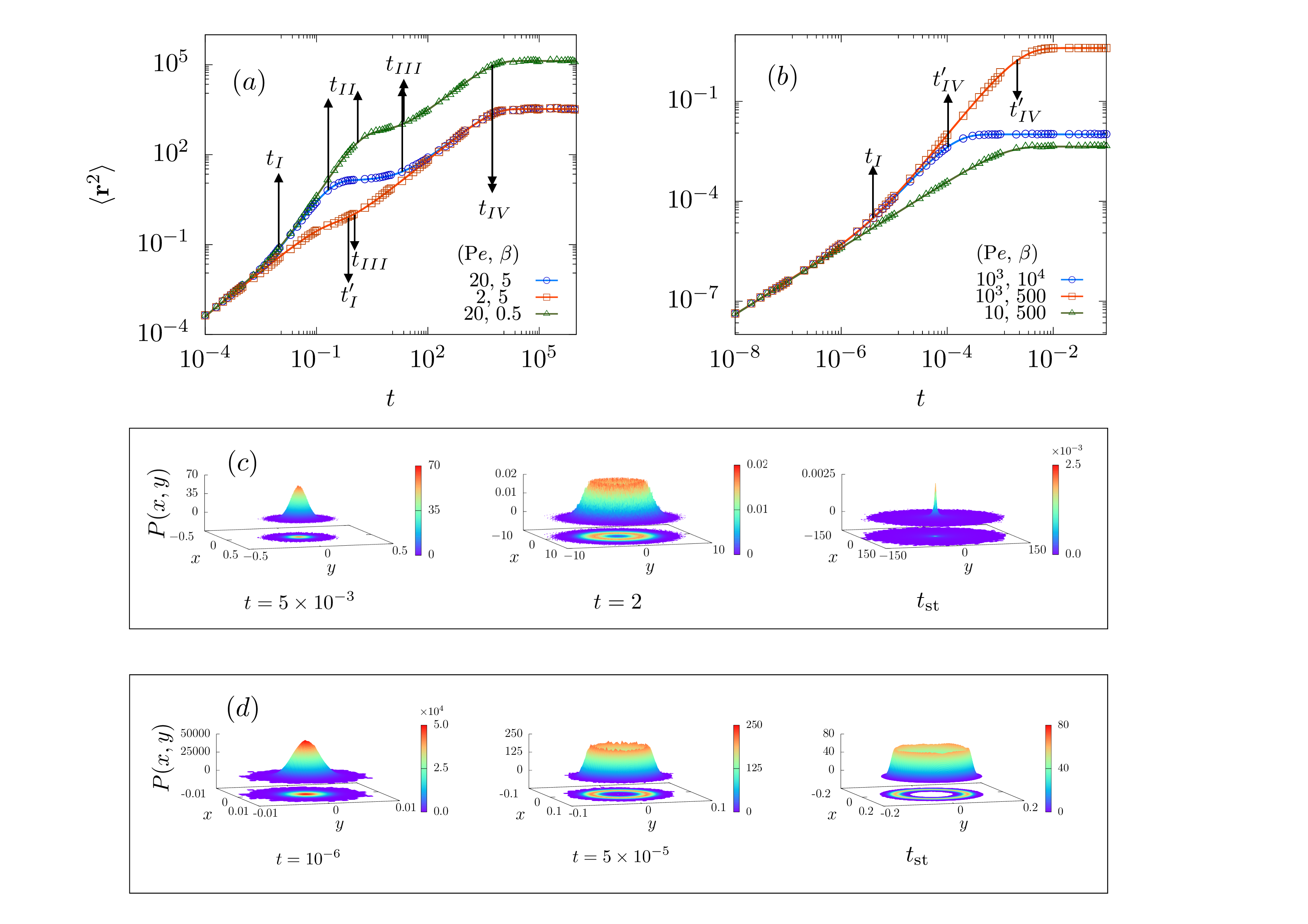}
    \caption{Time evolution of second moment of position $\la \rv^2 \ra $ and probability distribution $P(x,y)$ in two dimensions. ($a$) {\textit{Slow speed relaxation:}} Parameters used are $\t_v = 10^{4}$, $A_v = 10$ with $(\Pe, \,\beta )=(20,\,5) (\mathrm{blue~circle}, \circ)$, $(\Pe, \,\beta )=(2,\,5) (\mathrm{red~square}, \square)$, and $(\Pe, \,\beta )=(20,\,0.5) (\mathrm{green~triangle}, \triangle)$. ($b$) {\textit{Fast speed relaxation:}} Parameters used are $\t_v = 10^{-2}$, $A_v = 10$ with  $(\Pe, \,\beta )=(10^{3},\,10^{4}) (\mathrm{blue~circle}, \circ)$, $(\Pe, \,\beta )=(10^{3},\,5 \times 10^{2}) (\mathrm{red~square}, \square)$, and $(\Pe, \,\beta )=(10,\,5 \times 10^{2}) (\mathrm{green~triangle}, \triangle)$.
    The line plot is obtained using equation~(\ref{eq:r2d2}), and the points denote the simulation results.
    In ($a$) and ($b$), intial conditions are $\rv_0  = (0,0), \, \uv_0 = (1,0),\,v_1 = \Pe$. 
     In ($c$) and ($d$), we show the time evolution of probability distribution corresponding to the parameters of (blue circle, $\circ$) from ($a$) in $(c)$ and (blue circle, $\circ$) from ($b$) in ($d$), with the initial orientation chosen in all possible directions.
        }
    \label{fig:fig1}
\end{figure}

%%%%%%%%%%%%%

We illustrate the dynamical crossovers by plotting {a comparison between direct numerical simulation results (denoted by points) and analytical expression in} equation~(\ref{eq:r2d2}) {(lines)} in Figure~\ref{fig:fig1}($a$) for $\tau_v = 10^{4}$ and $A_v = 10$. Crossover times for various $\Pe$ and $\beta$ values are marked, showing good agreement with numerical results. For $\Pe = 20$ and $\beta = 5, \, 0.5$ (where $\Pe^2 > 2d \beta$), four crossovers occur at $t_{I}, \,t_{II}, \,t_{III},$ and $t_{IV}$. In contrast, for $\Pe = 2$ and $\beta = 5$ (where $\Pe^2 < 2d \beta$), three crossovers are observed at $t'_{I},\, t_{III}$, and $t_{IV}$.

Further, from direct numerical simulations, we obtain the probability distribution of position for $\Pe = 20$ and $\beta = 5$, with initial orientations in all directions, as shown in Figure~\ref{fig:fig1}($c$). Initially, the distribution is a small Gaussian due to dominant translational diffusion. At intermediate times, it forms a ring-like shape, indicating a quasi-steady state. 
As $\tau_v \gg 1$, the relaxation of active speed up to $t=2$ can be neglected, meaning the distribution is determined by the initial condition. Consequently, ABPs behave like constant-speed ABPs in a harmonic trap~\cite{Chaudhuri2021}, resulting in the metastable ring distribution when $\Pe^2 > \beta$. 
At steady state, the distribution becomes heavy-tailed, characterized by a positive excess kurtosis discussed in Section~\ref{sec:fourthmoment}. The full time-evolution of the analytically obtained kurtosis explains the deviation from Gaussian behavior, showing negative values at intermediate times and positive values later, as illustrated in Figure~\ref{fig:kurtosis time} in Appendix~\ref{appendix-c}.

Similar steady-state heavy-tailed probability distribution in position has been observed before for direction-reversing active Brownian particles (DRABPs) in a harmonic trap \cite{Santra21sm}. Our model can be mapped to DRABPs up to the second moment by setting the steady-state activity $\Pe = 0$. For DRABPs, the active velocity is given by ${\boldsymbol{v}_d} = v_d \sigma \uv$, with dichotomous noise $\sigma = \pm 1$ and $v_d$ representing the active speed's magnitude. 
In steady state, the active velocity satisfies $\la \boldsymbol{v}_{d} \ra = 0$ and has a correlation function $\la \boldsymbol{v}_{d}(t) \cdot \boldsymbol{v}_{d}(t') \ra = v_d^2 e^{-\gamma |t-t'|} e^{-(d-1)|t-t'|}$, where $\gamma$ controls the rate of change of the dichotomous noise. As $\Pe \to 0$, our fluctuating speed ABP model shows $\la \boldsymbol{v} \ra = 0$ and the correlation function $\la \boldsymbol{v}(t) \cdot \boldsymbol{v}(t') \ra = \tau_v A_v \, e^{-|t-t'|/\tau_v} e^{-(d-1)|t-t'|}$, resulting in the relations $v_d^2 = \tau_v A_v$ and $\gamma = 1/\tau_v$.

\subsection*{ (ii) Shallow trap strength and fast speed relaxation: $\t_v \ll t \ll \be^{-1}$}
In the limit of shallow trap strength and fast speed relaxation at intermediate times, $\t_v \ll t \ll \be^{-1}$, one can use $e^{-\be t} \approx 1 - \be t$, $e^{-(d-1+\be)t} \to 0$ and $e^{-t/\t_v} \to 0$. Thus, the series expansion of equation~(\ref{eq:r2d2}) at an intermediate time yields  $\la \rv^2 \ra_{\rm int } \approx {\cal C}+ {\cal D} t$, where
\begin{align}
\label{eq:cvar}{\cal C} &= - \left[\frac{2 \Pe^2}{ (d-1)^2 -\be^2}+\frac{A_v \t_v^3 [3 + (d-1- \be) \t_v]}{(1 - \be \t_v)(1-(d-1-\be) \t_v)(1+(d-1+\be)\t_v)} \right]\,, \\
{\cal D} &= \frac{2 [\be d (d-1-\be) + \be \Pe^2 - A_v]}{\be (d-1-\be)} + \frac{2 A_v}{\be (d-1)(1- \be \t_v)} + \frac{2 A_v}{(d-1) (d-1-\be) [1+(d-1-\be) \t_v]}\,.
\end{align}
The intermediate time expression $\la \rv^2 \ra_{\rm int } \approx {\cal C}+ {\cal D} t$ is suggestive of a quasi-steady state as before. However, 
it is evident from equation~(\ref{eq:cvar}) that ${\cal C} < 0$ in this parameter regime of $\be \t_v \ll 1$  making it impossible for ABPs to attain an intermediate time quasi-steady state. Thus, beyond the initial diffusive-ballistic crossover at $t_I$ discussed after equation~\eqref{eq:r2d_small_time}, the system can only show a subsequent ballistic-diffusive crossover at $t'_{III}={\cal D}/(\Pe^2-2d\be)$ provided the parameter values allow $t'_{III} > t_I$, before reaching the final steady state at $t_{\rm st} = \la \rv^2_{\rm st } \ra/{\cal D} $.

\subsection*{(iii) Strong trap strength and fast speed relaxation: }
In the limit of fast speed relaxation and strong trap strength $ [\beta^{-1}, \tau_v] \ll t $, both $ e^{-\beta t} $ and $ e^{-t/\tau_v} $ approach 0. The MSD reaches the steady-state value described by equation~(\ref{eq:r2d_steady_state}) directly, after the initial diffusive-ballistic crossover at $t_I$. The crossover from ballistic regime to steady state happens at  $ t_{IV}' = \left[\langle r^2 \rangle_{\text{st}} / (\Pe^2 - 2d\beta)\right]^{1/2} $, obtainable by comparing the ballistic term in equation~(\ref{eq:r2d_small_time}) with the steady state described by equation~(\ref{eq:r2d_steady_state}).

We plot the MSD in the fast speed relaxation regime in figure~\ref{fig:fig1}($b$) with parameters $ \tau_v = 0.01 $ and $ A_v = 10 $ to illustrate the crossovers. For $ \Pe = 10^{3} $ and $ \beta = 10^{4}, 500 $ (where $ \Pe^2 \gg 2 d \beta $), we observe two crossovers at $ t_I $ and $ t_{IV}' $. In contrast, with $ \Pe = 10 $ and $ \beta = 500 $ (where $ \Pe^2 < 2 d \beta $), there is a single crossover from diffusion to global trapping at $ \beta^{-1} $.
Note that at $ \Pe^2 < 2d\beta $, the above-mentioned behavior is equilibrium-like. In this parameter regime, as $ t_I <0$, no diffusive-ballistic crossover is possible, the initial diffusive dynamics directly reaches the final steady state; as in equilibrium ($\Pe=0$, $\tau_v=0$, $A_v=0$) this happens at $ \beta^{-1} $

For $ \Pe = 10^3 $ and $ \beta = 10^4 $, we analyze the time evolution of the position probability distribution in figure~\ref{fig:fig1}($d$), which begins as a Gaussian determined by translational diffusion. It then evolves into a ring-like distribution with the ring-size expanding until reaching a steady state defined by $ r \sim \Pe/\beta $. This ring-like distribution exhibits negative excess kurtosis. Thus the evolution is further elucidated by the excess kurtosis over time, as shown in figure~\ref{fig:kurtosis time} in Appendix~\ref{appendix-c}.
}

\section{Fourth Moment, kurtosis and phase diagram}
\label{sec:fourthmoment}
In this section, we explore the time evolution of the fourth moment of displacement $\la \rv^4 \ra (t)$ and present its exact form at the steady state. 
Further, we analyze the excess kurtosis using the fourth and second moments of displacement and present its exact form at the steady state. This allows us to comprehensively characterize the steady-state non-equilibrium features of the displacement distribution in terms of `phase diagrams' characterizing the distribution at different parameter regimes. 

\subsection{Fourth moment of displacement}
\begin{figure}
    \centering
    \includegraphics[width=17cm]{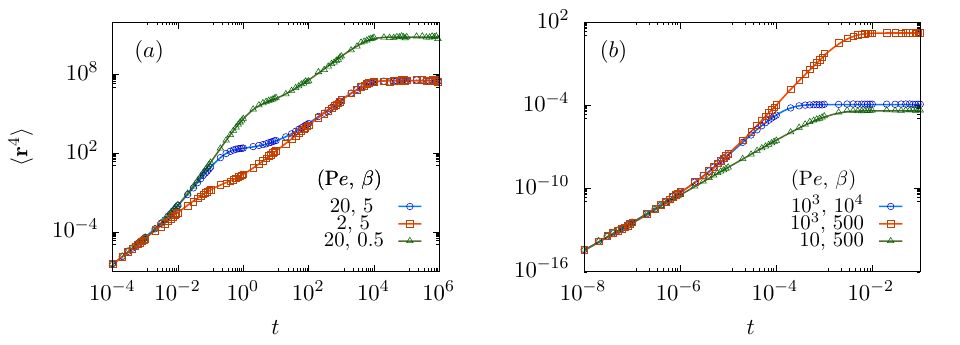}
    \caption{Time evolution of fourth moment of position $\la \rv^4 \ra (t)$ in $2d$. Parameter values, initial conditions, and line-point types used in ($a$) and ($b$) correspond to that of Fig.\ref{fig:fig1}($a$) and ($b$), respectively. Lines denote analytic results, and points denote numerical simulations. }  
%    ($a$)~{\textit{Slow speed relaxation:}} Parameters used are $\t_v = 10^{4}$, $A_v = 10$ with $(\Pe, \,\beta )=(20,\,5) (\mathrm{blue~circle}, \circ)$, $(\Pe, \,\beta )=(2,\,5) (\mathrm{red~square}, \square)$, and $(\Pe, \,\beta )=(20,\,0.5) (\mathrm{green~triangle}, \triangle)$. 
%    ($b$) {\textit{Fast speed relaxation:}} Parameters used are $\t_v = 10^{-2}$, $A_v = 10$ with  $(\Pe, \,\beta )=(10^{3},\,10^{4}) (\mathrm{blue~circle}, \circ)$, $(\Pe, \,\beta )=(10^{3},\,5 \times 10^{2}) (\mathrm{red~square}, \square)$, and $(\Pe, \,\beta )=(10,\,5 \times 10^{2}) (\mathrm{green~triangle}, \triangle)$. The initial choice of position $\rv_{0} = \bf{0}$ and $v_1 = \Pe$ was kept fixed. The line denotes the analytic estimates of $\la \rv^4 \ra$ obtained using equation~(\ref{eq:r4s}-\ref{eq:v3s}), and the points denote the simulation results.}
    \label{fig:r4}
\end{figure}

The calculation for $\la \rv^4 \ra(t)$  follows the same steps as shown for other moments in the previous sections and utilizes equation\,(\ref{eq:moment}).
The main steps to obtain the fourth moment of position in the Laplace space $\la \rv^4 \ra_s$ are shown in Appendix~\ref{appendix-c}.
One can solve equations\,(\ref{eq:r4s})-(\ref{eq:v3s}) to get $\la \rv^4 \ra_s$ and take the inverse Laplace transform for the full-time evolution $\la \rv^4 \ra (t)$. 
The full expression for $\la \rv^4 \ra (t)$ is prohibitively long to present here.
Therefore, using $d=2$, we plot the expression in figure~\ref{fig:r4}  and compare it with numerical simulation to obtain a good agreement.

In the following,  we analyze the behavior of this moment in short and long time limits. 
The series expansion of $\la \rv^4 \ra (t)$ around $t = 0$ for initial values $\rv_0 = 0$ and $v_1 = \Pe$ in $2d$ gives (see Appendix~\ref{appendix-d} for results in $d = 3$ dimensions)
\bea
\label{eq:r4ser2d}
 \la \rv^4 \ra (t) = 32 t^2 + 16 (\Pe^2 -4 \beta) t^3 + \left[ \frac{32}{3}(A_v + 7 \beta^2)-\frac{16}{3}\Pe^2(1+6 \beta) + \Pe^4 \right] t^4 + {\cal O}(t^5)\,.\nn\\
\eea
$\la \rv^4 \ra (t)$ shows a $t^2$ behavior at small time that crosses over to $t^3$ at $t_{I} = 2/(\Pe^2 - 4 \beta)$ dependent on the trap strength. The impact of speed fluctuation and relaxation in terms of $A_v$ and $\t_v$ appears at a much later time.  

In the long time limit, $\la \rv^4 \ra (t)$ reaches a steady state. The steady-state value in $2d$ (see Appendix~\ref{appendix-d} for $d = 3$ dimensions) is given by
\begin{align}
\label{eq:r4steady2d}
 \la \rv^4 \ra_{\rm st} &= \frac{8 \Pe^2}{\beta^2(1+\beta)} + \frac{{\cal A}_0 A_v \Pe^2 }{{\cal B}_0} + \frac{(4+3\beta) \Pe^4}{\beta^2(2 + 9 \beta + 10 \beta^2 + 3 \beta^3)} + \frac{8}{\beta^2} + \frac{8 \t_v^2 A_v}{\beta^2(1+\t_v + \beta \t_v)}  \nn\\
 &+ \frac{\t_v^4 A_v^2 [ 4 + 8 \t_v + \beta(3 +2(13 + 6 \beta)\t_v + 3(2+\beta)(4+3 \beta) \t_v^2)]}{\beta^2(2+\beta)(1+\beta \t_v)(1 + \t_v + \beta \t_v)(1+\t_v + 3 \beta \t_v)(1+(2+\beta)\t_v)}\,,
\end{align}
where ${\cal A}_0 = 2 \t_v^2 [4 + \beta(17 + 9 \beta) +20 \t_v + \beta \t_v (119 + \beta(172+63 \beta)) + 4 \t_v^2 (4 +\beta(53 + 4 \beta(37 + 33 \beta + 9 \beta^2))) + 12 \beta \t_v^3 (1+ \beta)(2 + \beta) (1+3 \beta) (4 + 3 \beta) ]$ and ${\cal B}_0 = \beta^2 (1+\beta)(2+ \beta)(1+3 \beta)(1+ \t_v + \beta \t_v) (1+2\t_v \beta)(1+ \t_v + 3 \t_v \beta)(1+2 \t_v (2 + \beta)) $.

In the limit of $t_v \to 0$ or $A_v \to 0$, the steady-state fourth moment recovers the result for constant speed ABPs in harmonic trap~\cite{Chaudhuri2021}
\bea
\la \rv^4 \ra_{\rm st} = \frac{8}{\be^2} + \frac{ \Pe^2}{\be^2 (1+\be)} \left( 8 + \frac{(4 + 3\be) \Pe^2}{(2+\be)(1+3 \be)} \right)\, . 
\eea
Notably,  $\la \rv^4 \ra_{\rm st}$ increases with increasing speed-relaxation time $\t_v$ or speed fluctuation strength $A_v$.  
On the other hand, it decreases with the trap stiffness $\be$, as expected.

\subsection{Kurtosis and `Phase diagram'}
\label{phase-diagram}
In this section, we present `phase diagrams' distinguishing the non-equilibrium distribution from a passive Gaussian process using the excess kurtosis. 
%We discuss the dynamics of excess kurtosis and dependency on the parameters in the steady state.  

The probability distribution of the displacement vector of a passive Brownian particle in the presence of a harmonic trap follows the Gaussian distribution. 
For such a process with $\la \rv \ra = 0$, the fourth moment of displacement is given by
\bea
\label{eq:mu4}
\mu_4 = \left(1 + \frac{2}{d} \right) \la \rv^2 \ra^2\,.
\eea
\begin{figure}
    \centering
    \includegraphics[width=17cm]{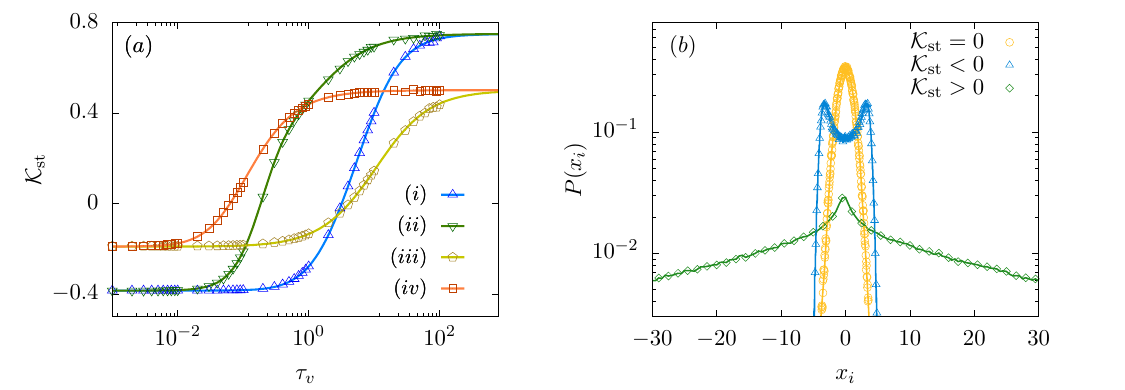}   
    \caption{Variation of excess kurtosis in $2d$ at steady state and the steady state probability distribution of position. $(a)$ The variation of ${\cal K}_{\rm st}$ as a function of $\t_v$ for the parameter values ($\be$, $\Pe$, $A_v$): ($i$) (1, 10, 10), ($ii$) (1, 10, $10^3$), ($iii$) (30, 10, 10), ($iv$) (30, 10, $10^3$). The solid lines represent the plot of equation~(\ref{eq:kur_st}), and points denote the simulation results. ($b$) The steady-state probability distribution $P(x_i)$ of position obtained from numerical simulation, where $x_i = x,y$. The parameter values ($\be$, $\Pe$, $A_v$, $\t_v$) for ${\cal K}_{\rm st} = 0$ are (2, 5, 1, 1); for ${\cal K}_{\rm st} < 0$ are (5, 20, 10, $10^{-2}$) and for ${\cal K}_{\rm st} > 0$ are (5, 20, 10, $10^4$). The points denote the simulation results, and the lines serve as guides for the eye.}
    \label{fig:line_kurtosis}
\end{figure}
The departure from this Gaussian behavior can be quantified in terms of the excess kurtosis, defined as
\bea
\label{eq:kur}
{\cal K} = \frac{\la \rv^4 \ra}{\mu_4} - 1\,,
\eea
where $\la \rv^4 \ra$ denotes the actual fourth moment of the displacement of the active process.

The exact time-dependence of the excess kurtosis can be calculated using the Laplace transform method we outlined before. However, the expression is prohibitively long, and we present the result plotting the time evolution in figure~\ref{fig:kurtosis time} in Appendix~\ref{appendix-c}. Note that the negative excess kurtosis corresponds to a bimodal distribution of marginal distribution along any direction $P(x)$ (corresponds to ring-like distribution in 2d, as shown in Fig.~\ref{fig:fig1}(d), for example) and positive excess kurtosis signifies a heavy-tailed distribution.

First, we analyze the excess kurtosis in the short and long time limits.
To get the short-time behavior, we series expand ${\cal K}(t)$ around $t = 0$ in $2d$ (see Appendix~\ref{appendix-d} for $d = 3$ dimensions)  
\begin{align}
\label{eq:kurtosis_ser2d}
 {\cal K}(t) &= \frac{-\Pe^4}{32}t^2 + \frac{(8 A_v \Pe^2 + 4 \Pe^4 + 3 \Pe^6)}{192} t^3 - \frac{1}{23040 \t_v} \left ( -320 A_v^2 \t_v + \Pe^4 \t_v \{ 136 -120 \beta^2 +45 \Pe^2 (8 + 3 \Pe^2)\} \right. \nn\\
&\left.  + 16 A_v \Pe^2 \{ 45 +\t_v(43-15 \beta + 15 \Pe^2)\}\right) t^4 + {\cal O}(t^5)\,,\nn\\
\end{align}
where we used initial values $v_1 = \Pe$ and $\rv = \bf{0}$.
Note that the initial deviation towards negative kurtosis is determined by $\Pe$. At a longer time, the negativity gets suppressed, aided by the speed fluctuation $A_v$. The impact of speed relaxation time $\tau_v$ and trapping strength $\beta$ impacts over a relatively longer time.

At the steady state, kurtosis takes the following form in $2d$ (see Appendix~\ref{appendix-d} for $d = 3$ dimensions)
\bea
\label{eq:kur_st}
{\cal K}_{\rm st} = \frac{{\cal A}_1 + {\cal A}_2 \Pe^2 + {\cal A}_3 \Pe^4}{{\cal B}_1}\,,
\label{eq_kst}
\eea
where
\begin{align}
 {\cal A}_1 &= \beta A_v^2 \t_v^4 (1+\beta)^2 (1+3 \beta) (1+2 \beta \t_v) [1+2(2+\beta) \t_v] [1+\t_v \{7 + 5 \beta+ (14 + \beta(20  + 7 \beta)) \t_v \nn\\
 &+ (2 + \beta) (8 + \beta(7+3 \beta)) \t_v^2  \}], \\
 {\cal A}_2 &= 2 \beta A_v \t_v^2 (1+\beta)(1+ \beta \t_v)(1+\t_v +\beta \t_v) (1+(2+\beta) \t_v) \left\{ 3 + 3\beta + 21 \t_v +\beta \t_v(44 +21 \beta) \right. \nn\\
 &\left. + 4 \t_v^2 [15 + 2 \beta(21 + 20 \beta + 6 \beta^2)] + 4 \t_v^3 (2+ \beta)(1+3 \beta) [8 + \beta(7+3 \beta)] \right\}, \\
 {\cal A}_3 &= -\beta(7+3 \beta) (1+\beta \t_v)(1+\t_v + \beta \t_v)^2 (1+ 2 \beta \t_v)(1+ \t_v + 3 \beta \t_v)[1+ \t_v (2+ \beta)] [ 1 + 2\t_v(2+\beta)], \\
  {\cal B}_1 &= 2 (2+\beta)(1+3 \beta)(1 + \beta \t_v) (1+2 \beta \t_v) (1 + \t_v + 3 \beta \t_v) (1+\t_v (2+\beta)) + (1+2\t_v(2+\beta)) \nn\\
  &[2 + 2\beta + \Pe^2 + \t_v (1+\beta) (2 + 2 \beta + \Pe^2) + A_v \t_v^2  (1+ \beta)]^2.
\end{align}
\begin{figure}
    \centering
    \includegraphics[width = 17cm]{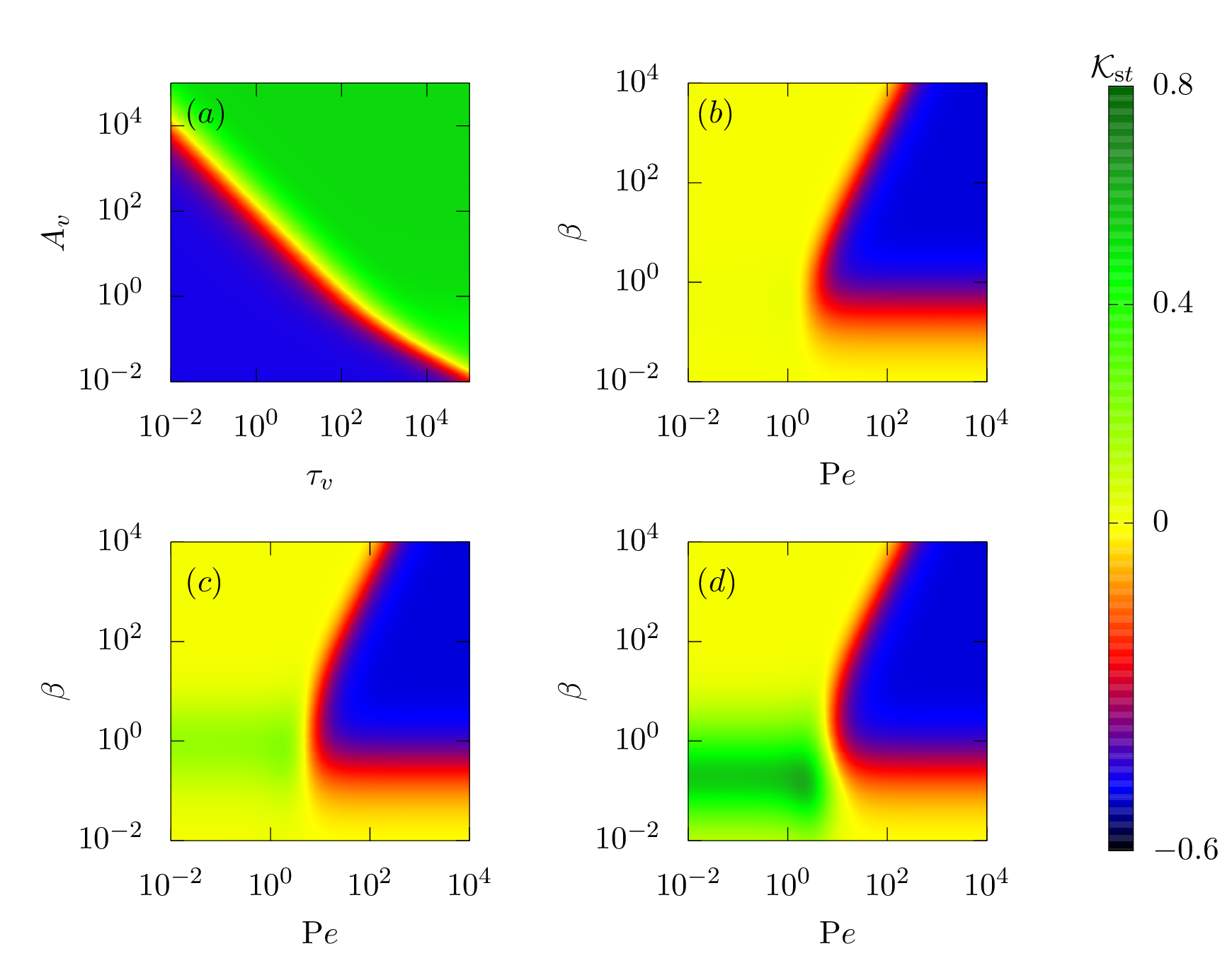}
    \caption{`Phase diagram': Heat maps of steady-state kurtosis ${\cal K}_{\rm st}$ in the $A_v - \t_v$ plane ($a$), and $\beta - \Pe$ plane ($b )- (d$) in $2d$. The parameter value for ($a$) are $\Pe = 20$, $\beta = 5$; for ($b$) are $\t_v = 1$, $A_v = 1$; for ($c$) are $\t_v = 1$, $A_v = 10$; and for $(d)$ are $\t_v = 10$, $A_v = 1$. The plots of kurtosis use equation~(\ref{eq:kur_st}).}
    \label{fig:kurtosis_heatmap}
\end{figure}

It is straightforward to analyze the behavior of ${\cal K}_{st}$ in the limits of small and large values of $\t_v$, $A_v$, $\be$, and $\Pe$.

{\it Limits in $\t_v$:} In the limit of $\t_v \to 0$, kurtosis restores the result of constant speed ABP in the harmonic trap and takes the form ${\cal K}_{\rm st} = -\beta \Pe^4 (7 + 3 \beta)/[2(2+ \beta)(1+3 \beta)(2 +2 \beta + \Pe^2)^2]$. 
In this regime, ${\cal K}_{\rm st} \leq 0$. In the other limit of $\t_v \to \infty$, it saturates to a positive value ${\cal K}_{\rm st} = (8 + 7 \beta + 3 \beta^2)/(4 +14 \beta + 6 \beta^2)$, which depends on $\be$ only such that $\lim_{\be \to 0} {\cal K}_{\rm st} = 2$ and $\lim_{\be \to \infty} {\cal K}_{\rm st} = 0.5$.

{\it Limits in $A_v$:}  With speed fluctuation, in the limit of $A_v \to 0$, kurtosis starts from ${\cal K}_{\rm st} =  -\beta \Pe^4 (7 + 3 \beta)/[2(2+ \beta)(1+3 \beta)(2 +2 \beta + \Pe^2)^2]$  in $2d$.
This value is equivalent to kurtosis for constant speed trapped ABPs \cite{Chaudhuri2021}. 
It increases with $A_v$ to saturate to a positive value in the limit of $A_v \to \infty$, which depends on $\beta$ and $\t_v$. The saturation can be understood by noting that ${\cal A}_1 \sim A_v^2$, ${\cal A}_2 \sim A_v$ while ${\cal A}_3$ is independent of $A_v$, and in the limit $A_v \to \infty$ ${\cal B}_1 \sim A_v^2$. 

{\it Limits in $\be$:}  As a function of $\be$, ${\cal K}_{\rm st}$ vanishes in two limits of large and small strength.
It vanishes as ${\cal K}_{\rm st} \sim \beta$ in the small $\beta$ limit. It varies non-monotonically at intermediate trap strength, eventually vanishing as ${\cal K}_{\rm st} \sim \beta^{-2}$ in the large trap strength limit. The limiting behavior of ${\cal K}_{\rm st}$ is the same as for simple ABP in a harmonic trap.

{\it Limits in $\Pe$:}   In the limit of $\Pe \to 0$, it can be easily seen from equation~(\ref{eq:kur_st}) that ${\cal K}_{\rm st} ={\cal A}_1/{\cal B}_1 > 0$, in the presence of speed fluctuation and relaxation. This contrasts the behavior of this system from traditional ABPs with constant active speed where ${\cal K}_{\rm st}$ vanishes as $\Pe \to 0$~\cite{Shee2020}. This is because of the active processes involved in the speed generation beyond $\Pe$.    
With an increase in $\Pe$, the excess kurtosis decreases and saturates to a negative value ${\cal K}_{\rm st} \approx - \beta(7 + 3 \beta)/(4 + 14 \beta + 6 \beta^2)$ in the limit of $\Pe \to \infty$ in $2d$. 
This negative value becomes maximum ${\cal K}_{\rm st} = -0.5$ for $\beta \to \infty$.

 We plot ${\cal K}_{st}$ using Eq.~\eqref{eq_kst} in figure~\ref{fig:line_kurtosis}($a$) as a function of relaxation time $\t_v$ at different $\Pe$, $\be$, $A_v$ values and find excellent agreement with numerical simulation results. All the graphs show a gradual increase of ${\cal K}_{st}$ with $\t_v$, starting from negative excess kurtosis corresponding to a bimodal marginal distribution of any component of displacement $P(x_i)$ to positive excess kurtosis corresponding to heavy-tailed distributions. Examples of such distributions obtained from direct numerical simulations are shown in figure~\ref{fig:line_kurtosis}($b$).

{\bf Phase diagram:}~
The kurtosis expression in equation~(\ref{eq:kur_st}) is used to obtain the heat maps in the $\t_v -A_v$, and $\Pe -\beta$ planes in $2d$, as shown in figure~\ref{fig:kurtosis_heatmap}. All the plots use the same color bar with blue (green) corresponding to regions of negative (positive) excess kurtosis to contrast them from the regions of equilibrium-like Gaussian behavior of vanishing kurtosis (yellow). 

At small $\t_v$ and $A_v$, the particle's dynamic behavior approaches that of simple ABPs, and in the presence of a reasonable trap stiffness and activity $(\Pe=20,\, \be=5)$ kurtosis takes a negative value, as shown in figure~\ref{fig:kurtosis_heatmap}($a$)~\cite{Chaudhuri2021}.
This behavior corresponds to the ring-like probability distribution of position at steady state, shown in figure~\ref{fig:fig1}($d$). 
With an increase in $\t_v$ or $A_v$, the effect of speed relaxation and fluctuation starts to dominate, and the value of excess kurtosis increases to take a positive value, shown in figure~\ref{fig:kurtosis_heatmap}($a$).
This corresponds to the late-time heavy-tailed distribution obtained in figure~\ref{fig:fig1}($c$) at steady-state.
The kurtosis variation from a negative to a positive value or vice-versa crosses by a zero kurtosis (equilibrium-like)  region.
 Notably, even in the limit $\t_v \gg 1$, a negative excess kurtosis is possible if $A_v$	
is sufficiently small, shown in figure~\ref{fig:kurtosis_heatmap}($a$).
 Likewise, a positive excess kurtosis can occur when $\t_v \ll 1$ and $A_v$ are sufficiently large, corresponding to a heavy-tailed distribution of positions.

At low values of $\t_v$ and $A_v$, the kurtosis phase diagram in the $\Pe-\be$ plane shown in figure~\ref{fig:kurtosis_heatmap}($b$) resembles that of constant speed ABPs in a harmonic trap~\cite{Chaudhuri2021}. This shows the possibility of re-entrant transition, e.g., by increasing $\be$ at a fixed $\Pe \approx 10$, the system transitions from vanishing (Gaussian distribution) to negative (bimodal distribution) to vanishing kurtosis. Similarly, around $\Pe=1$, as is visible from figure~\ref{fig:kurtosis_heatmap}($c$), getting a re-entrance from vanishing to positive (heavy-tailed distribution) to vanishing kurtosis is possible.  As $A_v$ or $\t_v$ increase, a region of positive kurtosis emerges and progressively expands, reducing the negative kurtosis region; see figures~\ref{fig:kurtosis_heatmap}($c$) and \ref{fig:kurtosis_heatmap}($d$). 
This highlights that slow-speed relaxation combined with large fluctuations leads to a positive excess kurtosis, while fast-speed relaxation with small fluctuation strength results in a negative excess kurtosis.

\section{Outlook} 
\label{sec_conc}
In this study, we presented and used an exact analytic method to explore the behavior of non-interacting active particles undergoing persistent random motion with fluctuating speeds in a $d$-dimensional harmonic trap in the presence of translational diffusion. This method enables the calculation of all time-dependent dynamical moments in any dimension. We presented explicit results for several moments, including second and fourth displacement moments, that closely match numerical simulations.

We examined dynamical crossovers in the mean squared displacement (MSD) under varying regimes of speed relaxation and trap strength, identifying crossover times between different scaling behaviors. Notably, under strong trap strength and slow speed relaxation, we observed a remarkable intermediate time metastability in the MSD, followed by a transition to a diffusive regime and eventually a steady state. This progression involves the evolution from an initial Gaussian distribution centered at the trap to an intermediate bimodal distribution with a spherically symmetric ring around the trap center, culminating in a heavy-tailed distribution peaking back at the center. The analytic results for excess kurtosis illustrate a non-monotonic change from zero (Gaussian) to negative values (bimodal) at intermediate times, then increasing to a positive value (heavy-tailed) in the steady state.

In steady state, swim pressure increases with activity as $\Pe^2$ and decreases with trap stiffness as $\beta^{-1}$. The effective diffusivity increases with activity $\Pe$, speed relaxation time $\tau_v$, and speed fluctuation strength $A_v$. We characterized the steady state using excess kurtosis, distinguishing non-equilibrium distributions from Gaussian ones, and presented detailed phase diagrams for four control parameters: speed relaxation time, speed fluctuation strength, activity, and trap strength. These diagrams reveal three distinct phases based on kurtosis: positive (heavy-tailed), negative (bimodal), and zero (Gaussian).

Our findings show that slow speed relaxation combined with large speed fluctuations favors positive excess kurtosis, while fast relaxation with low fluctuation strength corresponds to a simple active Brownian particle (ABP) and leads to negative kurtosis. The excess kurtosis also exhibited non-monotonic behavior with varying trap strength, showing possibilities of re-entrant transitions between Gaussian and non-Gaussian behaviors.

The analytical predictions from our study can be experimentally validated using artificial active particles, such as Janus colloids in harmonic traps. Additionally, our results can provide a foundation for understanding the dynamics of motile cells with speed and directional fluctuations in complex environments~\cite{Selmeczi2005, Frangipane2019, otte_2021}. In this paper, we employed Fokker-Planck equations to provide exact calculations for dynamical moments. However, deriving analytic expressions for the distribution functions presents an intriguing challenge that we believe would be a valuable direction for future research.

%In future work, a key objective will be to develop a comprehensive analytical framework for the probability distribution of speed-fluctuating ABPs. This probability distribution will help in further understanding the bimodal and heavy-tailed unimodal distributions observed in the excess kurtosis analysis. Work in this direction is underway.

%Outlook: Develop an analytical framework to describe the probability distributions. 
\section*{Acknowledgments}

The numerical calculations were supported in part by SAMKHYA, the high-performance computing facility at the Institute of Physics, Bhubaneswar. DC thanks Amit Singh Vishen for a critical reading of the manuscript, acknowledges a research grant from the Department of Atomic Energy (1603/2/2020/IoP/R\&D-II/15028), and thanks ICTS-TIFR, Bangalore, for an Associateship. AS acknowledges partial financial support from the John Templeton Foundation, Grant 62213.

\section*{Data availability statement}
All data that support the findings of this study are included in the article.

%\clearpage

\appendix

\section{Mean speed and speed fluctuation}
\label{appendix:mean-speed}
The mean speed can be easily calculated using equation~(\ref{eq:moment}) and is expressed~\cite{SheeJSTAT2022} as
\bea
\la v\ra(t)&=&v_1 e^{- t/t_v} + \Pe (1-e^{- t/\t_v}). 
\label{eq_v}
\eea
At the long time limit of $ t \to \infty$, it approaches the steady state value $\Pe$. Similarly, the second moment of the speed can be calculated as $\la v^2\ra(t)$~\cite{SheeJSTAT2022}
\bea
\la v^2 \ra(t)&=& \left[ v_1 e^{-t/\t_v} + \Pe\left( 1- e^{-t/\t_v} \right)\right]^2 
+\t_v A_v  \left( 1- e^{-2 t/\t_v} \right). 
\label{eq_v2}
\eea
Using equation~(\ref{eq_v}) and equation~\eqref{eq_v2},  we get the speed fluctuation, 
\bea
\la \d v^2\ra (t) &=& \la v^2\ra-\la v\ra^2=\t_v A_v\left(1-e^{-2t/\t_v}\right) .
\label{eq_dv2}
\eea
In the long time limit of $t \to \infty$, the equation gives the steady state fluctuations $\t_v A_v$. 

\section{Dependence of second moment on initial values}
\label{appendix:msdarb}
\begin{figure}
    \centering
    \includegraphics[width=17cm]{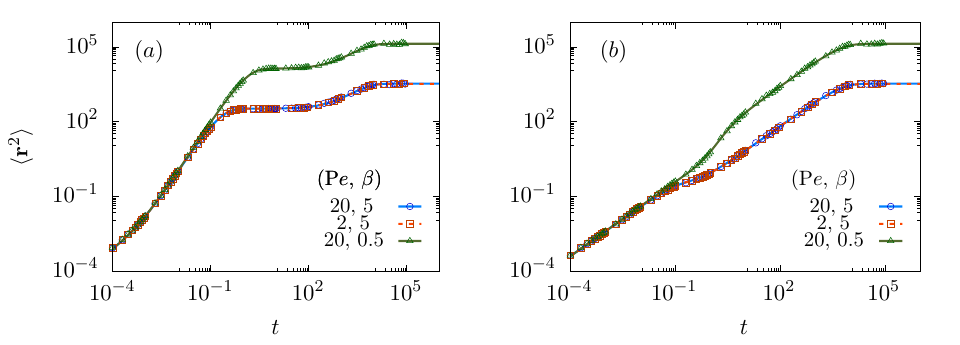}
    \caption{{Initial condition dependence of quasi-steady state:} Here we use parameter values as in Fig.\ref{fig:fig1} to show time evolution  $\la \rv^2 \ra (t)$ with two different initial conditions ($a$) $\rv_0  = (0.01,0), \, \uv_0 = (1,0),\,v_1 = 100 \neq \Pe$ and ($b$)  $\rv_0  = (0,0), \, \uv_0 = (1,0), \,v_1 = 0 \neq \Pe$, at $\t_v = 10^{4}$, $A_v = 10$ with $(\Pe, \,\beta )=(20,\,5) $(blue circle, $\circ$), $(\Pe, \,\beta )=(2,\,5)$ (red square, $\Box$), and $(\Pe, \,\beta )=(20,\,0.5)$ (green triangle, $\triangle$).  The points denote simulation results, and the lines are plots of analytic estimates obtained from equation~(\ref{eq:r2sf}).}
    \label{fig:r2_initial}
\end{figure}

For an arbitrary initial value of position $\rv_0$ and  active speed $v_1$, the MSD in Laplace space can be written as
\begin{align}
\label{eq:r2sf}
 \la \rv^2 \ra_s &= \frac{1}{s+2 \be} \left[  \rv_0^2 + \frac{2d}{s} + \frac{2}{s+(d-1)+\be + \t_v^{-1}} \left( \frac{2 \Pe^2 + 2 s\Pe \t_v v_1 +  \t_v (1+s\t_v) (2 A_v + s v_1^2)}{s(s \t_v + 1)(s \t_v + 2)}   \right. \right. \nn\\
& \left. \left. +v_1 \uv_0 \cdot \rv_0 + \frac{\Pe (\Pe + s(\uv_0 \cdot \rv_0 (1 + s \t_v) + \t_v v_1))}{s \t_v(s + d-1+\be)(s \t_v + 1)}   \right) \right].
\end{align}
The inverse Laplace transform of the above equation gives the full-time evolution of MSD.
However, as the expression is prohibitively lengthy to reproduce here, we plot the expression in figure~\ref{fig:r2_initial} and present its short time behavior analytically. 
The series expansion of MSD around $t=0$ gives
\bea
\label{eq:r2_ser_init}
\la \rv^2 \ra (t) &=& \rv_0^2 + 2 (d-\beta \rv_0^2 + v_1 \uv_0 \cdot \rv_0)t\nn\\
&+& \frac{[ \Pe \uv_0 \cdot \rv_0 - v_1 \uv_0 \cdot \rv_0 + \t_v \{ 2 \beta (-d + \beta \rv_0^2) + v_1^2 + v_1 \uv_0 \cdot \rv_0 - 3 (\beta +d) v_1 \uv_0 \cdot \rv_0 \}  ]}{\t_v} t^2 + {\cal O} (t^3).\nn\\
\eea
The comparison of the above expression with equation~(\ref{eq:r2d_small_time}) indicates that the impact of speed relaxation time $\t_v$ can be seen at an early time depending on the initial conditions.  
The initial conditions also result in a modification of the crossover times.
For the specific case where all particles start at the origin, equation~(\ref{eq:r2_ser_init}) simplifies to $\la \rv^2 \ra (t) = 2 d t + (v_1^2 - 2 d \be) t^2 + {\cal O}(t^3)$, showing that the intermediate time quasi-steady state depends on the initial active speed not the value of  $\Pe$.
To further emphasize the dependence on the initial active speed, we plot the MSD for the initial conditions $\rv_0 = (0.01, 0)$, $\uv = (1, 0)$, $v_1 = 100 \neq \Pe$, and $\rv_0 = (0, 0)$, $\uv = (1, 0)$, $v_1 = 0 \neq \Pe$ in the slow-speed relaxation regime in figure~\ref{fig:r2_initial}.
In figure~\ref{fig:r2_initial}($a$), where $v_1^2 > 2 d \beta$, a pronounced quasi-steady state characterized by intermediate time plateau of the MSD is observed at intermediate times. 
In figure~\ref{fig:r2_initial}($b$), where $v_1^2 < 2 d \be$, no such plateauing appear.
At the final steady-state, however, the observables are independent of the initial conditions, and the MSD is completely described by equation~(\ref{eq:r2d_steady_state}).

\section{Fourth moment calculation and Kurtosis evolution}
\label{appendix-c}
 To complete the calculation for the fourth moment, we utilize equation~(\ref{eq:moment}) to get 
\begin{align}
\label{eq:r4s} (s + 4 \beta) \la \rv^4 \ra_s &= \rv_0^4 + 4 (d+2) \la \rv^2 \ra_s + 4 \la v(\uv \cdot \rv) \rv^2 \ra_s, \\
 [s+(d-1)+3 \beta+1/\t_v] \la v(\uv \cdot \rv) \rv^2 \ra_s &=  v_1(\uv_0 \cdot \rv_0) \rv_0^2 + 2(d+2) \la v(\uv \cdot \rv) \ra_s + \la v^2 \rv^2 \ra_s + 2 \la v^2 (\uv \cdot \rv)^2 \ra_s 
 \nn\\ &+ \Pe \la (\uv \cdot \rv) \rv^2 \ra_s /\t_v, \\
 [s+2\beta + 2/\t_v] \la v^2 \rv^2 \ra_s &= v_0^2 \rv_0^2 + 2d \la \vv^2 \ra_s + 2 A_v \la \rv^2 \ra_s + 2 \la v^3 (\uv \cdot \rv) \ra_s + 2 \Pe \la v \rv^2 \ra_s /\t_v, \\
 [s+2d+2 \beta + 2/\t_v] \la v^2 (\uv \cdot \rv)^2 \ra_s &=  v_1^2 (\uv_0 \cdot \rv_0)^2 + 2 \la v^2 \ra_s + 2 \la v^2 \rv^2 \ra_s + 2 A_v \la (\uv \cdot \rv)^2 \ra_s 
 + 2 \la v^3 (\uv \cdot \rv ) \ra_s \nn\\ 
 &+ 2 \Pe \la v(\uv \cdot \rv)^2 \ra_s /\t_v, \\
 [s+(d-1)+3 \beta ] \la (\uv \cdot \rv) \rv^2 \ra_s &= (\uv_0 \cdot \rv_0) \rv_0^2 + 2(d+2) \la \uv \cdot \rv \ra_s + \la v \rv^2 \ra_s + 2 \la v(\uv \cdot \rv)^2\ra_s,  \\
 [s+(d-1)+ \beta + 3/\t_v] \la v^3 (\uv \cdot \rv) \ra_s &= v_1^3 (\uv_0 \cdot \rv_0) + 6 A_v \la v(\uv \cdot \rv)\ra_s + \la v^4 \ra_s + 3 \Pe \la v^2 (\uv \cdot \rv) \ra_s /\t_v, \\
 [s+2 \beta + 1/\t_v] \la v \rv^2 \ra_s &= 2d \la v \ra_s + 2 \la v^2 (\uv \cdot \rv) \ra_s + \Pe \la \rv^2 \ra_s/\t_v, \\
 [s+2d + 2\beta] \la (\uv \cdot \rv)^2 \ra_s &= (\uv_0 \cdot \rv_0)^2 + 2/s + 2 \la \rv^2 \ra_s + 2 \la v(\uv \cdot \rv) \ra_s, \\
 [s+2d + 2\beta + 1/\t_v] \la v (\uv \cdot \rv)^2 \ra_s &= v_1(\uv_0 \cdot \rv_0)^2 + 2 \la v \ra_s + 2 \la v \rv^2 \ra_s + 2 \la v^2 (\uv \cdot \rv) \ra_s + \Pe \la (\uv \cdot \rv)^2 \ra_s /\t_v, \\
 [s+4/\t_v] \la v^4 \ra_s &= v_1^4 + 12 A_v \la v^2 \ra_s + 4 \Pe \la v^3 \ra_s/\t_v,\\
 [s+(d-1)+ \beta + 2/\t_v] \la v^2 (\uv \cdot \rv) \ra_s &= v_1^2(\uv_0 \cdot \rv_0) + 2 A_v \la \uv \cdot \rv \ra_s + \la v^3 \ra_s  + 2 \Pe \la v(\uv \cdot \rv) \ra_s /\t_v, \\
\label{eq:v3s}  [s+3/\t_v] \la v^3 \ra_s &= v_1^3 + 6 A_v \la v \ra_s + 3 \Pe \la v^2 \ra_s/\t_v,
\end{align}
where $\la v \ra_s$, $\la v^2 \ra_s$, $\la \uv \cdot \rv \ra_s$, $\la v(\uv \cdot \rv) \ra_s$, and $\la \rv^2 \ra_s$ are already calculated in the main text.
Substituting the required moments in equation~(\ref{eq:r4s}), one gets the fourth moment of displacement in Laplace space, and taking the inverse Laplace transform, one can get the full time evolution. 

\begin{figure}
    \centering
    \includegraphics[width=10cm]{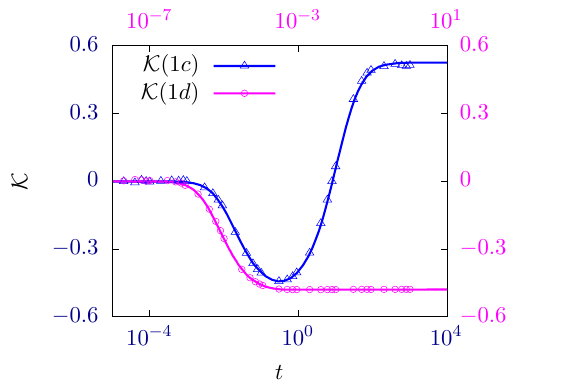}
    \caption{Time evolution of kurtosis corresponding to the parameters in figures~\ref{fig:fig1}($c$) and \ref{fig:fig1}($d$) in $2d$. The parameter values are $\Pe = 20$, $\beta = 5$, $\t_v = 10^4$, $A_v = 10$ for ${\cal K} (1c)$ and $\Pe = 10^3$, $\be = 10^4$, $\t_v = 10^{-2}$, $A_v = 10$ for ${\cal K}(1d)$. The line represents the analytic estimate, and the points denote the simulation results. Here, the dimensionless time $t$ is shown on a logarithmic scale.}
    \label{fig:kurtosis time}
\end{figure}

{Time evoltion of the fourth moment can be further utilized to derive the evolution of excess kurtosis ${\cal K}$, as described in equation~(\ref{eq:kur}). }The resulting expression is prohibitively long to present here. Thus, we show its time evolution in Fig.~\ref{fig:kurtosis time}, plotting its behavior for a couple of parameter values corresponding to Fig.~\ref{fig:fig1}($c$) and \ref{fig:fig1}($d$).   

\section{Kurtosis and fourth moments in three dimensions}
\label{appendix-d}
Solving equations~(\ref{eq:r4s}) - (\ref{eq:v3s}) and taking the inverse Laplace transform gives the complete time evolution of $\la \rv^4 \ra (t)$. To analyze initial time dependence, we expand $\la \rv^4 \ra(t)$ around $t=0$ in $d = 3$ dimensions to get 
\bea
\label{eq:r4exp3d}  \la \rv^4 \ra (t) = 60 t^2 + 20 (\Pe^2 - 6 \beta) t^3 +  \left [ 140 \beta^2 + \frac{40}{3} \left( A_v - (1+3 \beta) \Pe^2\right) + \Pe^4 \right] t^4 +{\cal O} (t^5). 
\eea
A comparison of the above expression with equation~(\ref{eq:r4ser2d}) indicates that the qualitative nature of crossovers seen in $\la \rv^4 \ra$ in $ d = 3$ dimensions is the same as in $2d$; however, the crossover time itself depends on the embedding dimension.

In the long time limit, the steady state value of $\la \rv^4 \ra$ in $d =3$ dimensions is given by
  \begin{align}
  \la \rv^4 \ra_{\rm st} &= \frac{10 \Pe^2}{\beta^2(2+\beta)} + \frac{{\cal A}^{'} A_v \Pe^2}{{\cal B}^{'} } + \frac{(5 + 3 \beta) \Pe^4}{\beta^2 (2+ \beta)(3+ \beta)(2 + 3 \beta)}+ \frac{15}{\beta^2} + \frac{10 A_v \t_v^2}{\beta^2(1+2 \t_v + \beta \t_v)} \nn\\
  & + \frac{\t_v^4 A_v^2 [5 + 15 \t_v + \beta(3 + \t_v(37 + 12 \beta) + 3(3 + \beta)(5 + 3 \beta) \t_v^2)]}{\beta^2(3 + \beta)(1+\beta \t_v)(1+(2+\beta) \t_v)(1+(3+\beta)\t_v)(1+(2+3 \beta)\t_v)},
 \end{align}
where ${\cal A}^{'} = 2 \t_v^2 [10 + \beta (25 + 9 \beta) + 80 \t_v + \beta \t_v(290 + \beta(263 + 63 \beta)) + 2 \t_v^2 (60 + \beta(448 + \beta (737 + 411 \beta + 72 \beta^2))) + 12 \beta \t_v^3 (2+\beta) (3 + \beta)(2 + 3 \beta)(5 + 3 \beta) ]$ and ${\cal B}^{'} = \beta^2 (2 + \beta) (3 + \beta) (2 + 3 \beta) (1 + 2 \beta \t_v) (1 + (2+ \beta) \t_v)(1+2(3 + \beta) \t_v) (1+(2+3 \beta) \t_v)$.

Following the definition of kurtosis from  equations~(\ref{eq:mu4})  and (\ref{eq:kur}), one can write the series expansion of kurtosis around $t = 0$ in  $d = 3$ dimensions as
\begin{align}
 {\cal K} (t) &= - \frac{\Pe^4}{90} t^2 + \frac{(8 A_v \Pe^2 + 4 \Pe^4 + \Pe^6)}{270} t^3 - \frac{1}{16200 \t_v} \left( -160 A_v^2 \t_v + \t_v \Pe^4 (152 - 30 \beta^2 + 15 \Pe^2 (8 + \Pe^2))   \right. \nn\\
& \left.  + 8 A_v \Pe^2 [45 + (82 - 15 \beta + 15 \Pe^2) \t_v] \right) t^4 + {\cal O} (t^5), \nn\\  
\end{align}
where the nature of crossover is the same as in $2d$.
Further, the steady-state value of kurtosis in $d = 3$ dimensions is given by
\bea
{\cal K}_{\rm st} = \frac{{\cal A}_4 + {\cal A}_5 \Pe^2 + {\cal A}_6 \Pe^4}{{\cal B}_3},
\eea
where
\begin{align}
 {\cal A}_4 &= 4 \beta A_v^2 \t_v^4 (2+ \beta)^2 (2+3\beta)  (1+2\beta \t_v)[1+2(3+\beta) \t_v] \left\{ 1 + \t_v [11+5\beta +(33  \right. \nn\\
 &\left. + \beta(31 + 7 \beta)) \t_v  + (3 + \beta) (15 + \beta(11 + 3 \beta)) \t_v^2]\right\}, \\
  {\cal A}_5 &= 8 \beta A_v \t_v^2  (2 + \beta) (1 + \beta \t_v) (1+(2+\beta) \t_v)(1+(3+\beta) \t_v) \left\{ 5 + 3 \beta + [55 + \beta(71   \right. \nn\\
 &\left.+ 21 \beta)] \t_v  + 2[111 + \beta (209 + \beta (127 + 24 \beta))] \t_v^2 + 4(3+\beta)(2+3 \beta)(15 + \beta(11+3 \beta)) \t_v^3 \right\}, \\
 {\cal A}_6 &= -2 \beta (11+3 \beta) (1+ \beta \t_v) (1 + 2 \beta \t_v)(1+(2+\beta) \t_v)^2 (1+(3 + \beta) \t_v) (1+2(3 + \beta) \t_v) \nn\\
 &(1+(2 + 3 \beta) \t_v), \\
 {\cal B}_3 &= 5(3 + \beta) (2+3 \beta) (1+ \beta \t_v) (1+ 2 \beta \t_v) (1+(3 + \beta) \t_v) (1+2(3+\beta) \t_v) (1+(2+3 \beta) \t_v) \nn\\
 &\left[ 6 + 3 \beta + \Pe^2 + (2+\beta)(6 + 3 \beta + \Pe^2) \t_v + (2+\beta) A_v \t_v^2 \right]^2.
\end{align}
The steady-state excess kurtosis in $d = 3$ exhibits the same three distinct `phases' as observed for $2d$ in the main text.
However, the phase boundaries depend on the embedding dimensionality of the system and is different in $d=3$ dimensions from $d=2$.
\section*{References}
%\bibliography{reference} 

\end{document}